\documentclass[12pt, letter, oneside]{article}
\usepackage{setspace}
\usepackage{longtable}
\usepackage{geometry}
\usepackage{graphicx}
\usepackage[parfill]{parskip}
\usepackage{booktabs}
\usepackage{graphics}
\usepackage{fullpage}
\usepackage{amsmath, mathtools, amssymb, amsthm}
\usepackage{multirow, multicol}
\usepackage{fancyhdr}
\usepackage{framed}
\usepackage{rotating}
\usepackage{epstopdf}
\usepackage[colorlinks, citecolor=blue]{hyperref}
\usepackage{bm}
\usepackage{xcolor}
\usepackage{listings}
\usepackage{natbib}
\usepackage{enumitem}
\usepackage[T1]{fontenc}
\usepackage{authblk}

\title{Statistical approaches using longitudinal biomarkers for disease early detection: A comparison of methodologies}
\author[1]{Yongli Han*}
\author[1]{Paul S. Albert}
\author[2]{Christine D. Berg}
\author[3]{Nicolas Wentzensen}
\author[1]{Hormuzd A. Katki}
\author[1]{Danping Liu}

\affil[1]{\small Biostatistics Branch, Division of Cancer Epidemiology \& Genetics, National Cancer Institute}
\affil[2]{\small Division of Cancer Epidemiology \&  Genetics, National Cancer Institute, MD, U.S.A.}
\affil[3]{\small Clinical Genetics Branch, Division of Cancer Epidemiology \&  Genetics, National Cancer Institute}

\date{}

\begin{document}
\maketitle
\footnote{*Correspondence: Yongli Han, Biostatistics Branch, Division of Cancer Epidemiology and Genetics, National Cancer Institute, MD, USA 20850, yongli.han@nih.gov}

\vspace{-3em}

\begin{abstract}
Early detection of clinical outcomes such as cancer may be predicted based on longitudinal biomarker measurements. Tracking longitudinal biomarkers as a way to identify early disease onset may help to reduce mortality from diseases like ovarian cancer that are more treatable if detected early. Two general frameworks for disease risk prediction, the shared random effects model (SREM) and the pattern mixture model (PMM) could be used to assess longitudinal biomarkers on disease early detection. In this paper, we studied the predictive performances of SREM and PMM on disease early detection through an application to ovarian cancer, where early detection using the risk of ovarian cancer algorithm (ROCA) has been evaluated. Comparisons of the above three methods were performed via the analyses of the ovarian cancer data from the Prostate, Lung, Colorectal, and Ovarian (PLCO) Cancer Screening Trial and extensive simulation studies. The time-dependent receiving operating characteristic (ROC) curve and its area (AUC) were used to evaluate the prediction accuracy. The out-of-sample predictive performance was calculated using leave-one-out cross-validation (LOOCV), aiming to minimize the problem of model over-fitting. A careful analysis of the use of the biomarker cancer antigen 125 for ovarian cancer early detection showed improved performance of PMM as compared with SREM and ROCA. More generally, simulation studies showed that PMM outperforms ROCA unless biomarkers are taken at very frequent screening settings. 

Keywords: Disease early detection prediction; pattern mixture model; risk of ovarian cancer algorithm; shared random effects model; time-dependent AUC
\end{abstract}

\section{Introduction}\label{sec1}
Epidemiologic studies usually incorporate longitudinal biomarkers into the prediction of clinical outcomes. Early disease detection could also benefit from this approach, since additional information on critical time point and pathology is often contained in the subject-specific biomarker trajectories \citep{drescher2013longitudinal, menon2015risk, han2019accounting}. Tracking longitudinal biomarkers in a population may result in earlier disease detection and may help to reduce mortality from diseases that are more treatable if detected early \citep{mcintosh2003parametric}. One example is ovarian cancer, which is the fifth leading cause of cancer-related deaths among the U.S. women and one of the most lethal gynecological cancers \citep{skates2003calculation, zhang2004three, clarke2009screening, henderson2018screening, russell2017novel}. Ovarian cancer usually has no symptoms at its early stage and develops undetected until it has spread within the pelvis and abdomen \citep{matulonis2016ovarian}. 
The U.S. ovarian cancer survival statistics show that the 5-year survival rate for women with late stage ovarian cancer is only 29.2\% (23\% in the U.K.), in contrast to 92.4\% (about 90\% in the U.K.) for those diagnosed at an early stage \citep{howlader2019seer, russell2017novel}. Although early stage ovarian cancer can be treated with a higher success rate \citep{clarke2009screening}, the majority of ovarian cancer cases are diagnosed at  late stage, when curative treatment rarely exists, making methods for the detection of early stage ovarian cancer desirable \citep{skates2017early}. However, large randomized trials have not shown a survival benefit for current early detection approaches of ovarian cancer so far \citep{pinsky2013potential, wentzensen2016large}.

When the interest is to use longitudinal biomarker information to predict a subsequent binary outcome, good modeling of the longitudinal biomarker trajectories is often the key to obtain accurate outcome prediction. However, in the prediction of ovarian cancer early detection, there is not much research to investigate how the different strategies of modeling the biomarker trajectories would affect the prediction accuracy under the longitudinal setting. Besides, a comparison of the different techniques is complicated, in part because it may depend on the screening frequency of the biomarker.

The risk of ovarian cancer algorithm (ROCA) has been proposed for  ovarian cancer early detection using repeatedly measured serum biomarker cancer antigen 125 (CA-125) values \citep{skates2001screening}. Specifically in the model setting, ROCA separately models the longitudinal CA-125 trajectories for the cases and the controls. For the controls, a constant mean model of CA-125 is assumed with a random intercept term that accounts for subject heterogeneity. For the cases, the CA-125 trajectory is assumed to be piecewise linear with a latent subject-specific changepoint. The probability of early detection is then constructed using Bayes' theorem. 

Recently, two general frameworks for disease risk prediction by modeling longitudinal biomarker behaviors have been developed. The shared random effects model (SREM) \citep{albert2012linear} predicts a binary outcome based on the longitudinal biomarkers by assuming a shared random effects structure that links the binary outcome and the longitudinal process together, while the pattern mixture model (PMM) \citep{liu2014combination} fits the biomarker distributions conditional on the binary outcome and then constructs the outcome prediction using Bayes' theorem. SREM and PMM are originally proposed for disease risk prediction using serial biomarker values and associated observation times \citep{han2019accounting, liu2014combination}, but can be applied more generally to predict disease early detection in the longitudinal setting. 

In this paper, we focused on examining and evaluating the utility of SREM and PMM for disease early detection through an application to ovarian cancer, under which SREM and PMM were compared with ROCA. Comparisons of SREM and PMM with ROCA were performed via simulation studies and an empirical analysis of the ovarian cancer data from the Prostate, Lung, Colorectal and Ovarian (PLCO) Cancer Screening Trial. Specifically, we first extended ROCA to identify the potential effects of the baseline age and the screening time on the marker trajectories and then proposed a maximum-likelihood approach for parameter estimation. Specific formulations of SREM and PMM for predicting the early detection of ovarian cancer were also proposed. Modeling forms under SREM, PMM, and ROCA for the longitudinal CA-125 trajectories in the PLCO Cancer Screening Trial were compared and their effects on the prediction accuracy of ovarian cancer early detection were further assessed. The predictive performances of different models were evaluated by the time-dependent receiving operating characteristic (ROC) curve and its area (AUC) \citep{heagerty2000time}, such that the right censored cancer diagnosis times can be incorporated into the AUC calculation. Previous studies, to the contrary, simply used an ordinary ROC curve to examine the accuracy of ROCA \citep{russell2017novel} by treating the ovarian cancer outcome as binary and hence ignored the diagnosis time information. Furthermore, to estimate the out-of-sample prediction accuracy, we applied the leave-one-out cross-validation (LOOCV) technique to minimize the problem of model overfitting \citep{russell2017novel, berchuck2005patterns}. In addition, to check the effects of biomarker measuring frequency on the prediction accuracy of SREM, PMM, and ROCA regarding ovarian cancer early detection, we considered three screening frequencies (annual, biannual, and quarterly) in the simulation studies. Our research answered the questions about the effects of using SREM, PMM, and ROCA as well as the effects of different screening frequencies on the prediction accuracy of ovarian cancer early detection.

The rest of this paper is organized as follow: Section \ref{section1} briefly reviews the general settings of the SREM and PMM frameworks. Section \ref{methods} first introduces the problem of predicting the ovarian cancer early detection and reviews ROCA. Potential issues of ROCA are pointed out and several extensions are then proposed. In the end, the formulations of SREM and PMM tailored for predicting ovarian cancer early detection are specified. We compare the prediction performances of PMM, SREM, and ROCA in Section \ref{plco} through an application to the PLCO ovarian cancer trial and perform additional simulation studies in Section \ref{simulation}. Section \ref{discussion} ends this study with a discussion.

\section{Review of Methods}\label{section1}
In this section, we review the SREM and the PMM frameworks. Without loss of generality, let $Y_{ij}$ denote the biomarker value for the $i$th individual at time $t_{ij}$, where $i=1, 2, \ldots, N$, $N$ is sample size, $j=1, 2, \ldots, n_{i}$, and $n_{i}$ is the number of longitudinal biomarker measurements for the $i$th participant. Without loss of generality, assume the first $M$ subjects are cases and the rest $N-M$ are controls. Let $D_{i}$ be the binary outcome, such that $D_{i}=1$ indicates a case and $D_{i}=0$ denotes a control.  
\subsection{Shared Random Effects Model (SREM)} \label{SREM}
SREM jointly models the longitudinal biomarker trajectories and the binary disease outcomes \citep{albert2012linear}, which are assumed to share the same set of random effects. For the case and the control trajectories, a linear mixed model was proposed
\begin{align}
\bm{Y}_{i} = \bm{X}_{i}\bm{\theta}+\bm{Z}_{i}\bm{b}_{i} +\bm{\xi}_{i} \label{srem1}
\end{align}
where $\bm{X}_{i}$ and $\bm{Z}_{i}$ are design matrices of the fixed and the random covariates, $\bm{\theta}$ are the fixed effects, $\bm{b}_{i}$ stand for the random effects, and $\bm{\xi}_{i}\sim MVN(\bm{0}, \bm{\Sigma}_{\xi})$ are the random measurement errors. The relation between $D_{i}$ and $\bm{b}_{i}$ is given as
\begin{align}
P(D_{i}=1|\bm{b}_{i}) = g\left\{\alpha_{0}+\bm{\alpha}_{1}^{\top}h(\bm{b}_{i})\right\} \label{link1}
\end{align}
where $g(\cdot)$ is a link function and $h(\cdot)$ is a function of random effects. In the simulation studies and the real data analyses, we set $\bm{\alpha}_{1}^{\top}h(\bm{b}_{i})$ to be $\alpha_{1}b_{i0}+\alpha_{2}b_{i1}+\alpha_{3}b_{i2}+\alpha_{4}b_{i3}$, so each random effect would affect the disease outcome differently. The strength of association between the longitudinal process and the binary outcome is quantified by $\bm{\alpha}_{1}$, while $\alpha_{0}$ is the intercept. 

As SREM gives the distributions of $\bm{Y}_{i} |\bm{b}_{i}$ and $D_{i}=1|\bm{b}_{i}$, the joint distribution of $\bm{Y}_{i}$ and $D_{i}$ can be derived by integrating over the random effects $\bm{b}_{i}$. The diagnosis probability thus can be calculated from the joint distribution of $Y_{i}$ and $D_{i}$ as
\begin{align}
P(D_{i}=1|\bm{Y}_{i})  = \frac{\int P(\bm{Y}_{i}|\bm{b}_{i})P(D_{i}=1|\bm{b}_{i})f(\bm{b}_{i})\text{d}\bm{b}_{i}}{\int P(\bm{Y}_{i}|\bm{b}_{i})f(\bm{b}_{i})\text{d}\bm{b}_{i}} \label{srem-risk} 
\end{align}
where $f(\bm{b}_{i})$ is the probability density function (PDF) of $\bm{b}_{i}$. If a probit link function is used and $\bm{b}_{i}\sim MVN(\bm{0}, \bm{\Sigma}_{\bm{b}})$, \cite{albert2012linear} revealed that the SREM estimation can be substantially simplified by decomposing the joint likelihood function of $D_{i}$ and $\bm{Y}_{i}$ as
\begin{align}
L = L_{1}\times L_{2} = \prod_{i=1}^{N}f(\bm{Y}_{i})\prod_{i=1}^{N}f(D_{i}|\bm{Y}_{i}) \label{srem-like}
\end{align}
where $L_{1}$ is from the longitudinal process in (\ref{srem1}) and $L_{2}$ has an explicit expression 
\begin{align}
P(D_{i}=1|\bm{Y}_{i}) = \Phi\left\{\frac{\alpha_{0}+\bm{\alpha}_{1}h(\bm{b}_{i})}{\sqrt{1 + \bm{\alpha}_{1}^{\top}Cov(h(\bm{b}_{i}))\bm{\alpha}_{1}}} \right\} \label{srem-risk1}
\end{align}
Parameters in (\ref{srem1}) are estimated by maximizing $L_{1}$, while $\alpha_{0}$ and $\bm{\alpha}_{1}$ in (\ref{link1}) are estimated by maximizing $L_{2}$, given the estimates from $L_{1}$. The probability is then obtained by replacing the parameters in (\ref{srem-risk1}) with their maximum likelihood estimates (MLEs). 
\subsection{Pattern Mixture Model (PMM)} \label{PMM}
PMM directly makes the assumption of a linear mixed model on the biomarker trajectories conditional on the binary outcome, i.e., $\bm{Y}_{i}|D_{i}=d, d=0, 1$. In other words, PMM formulates the longitudinal behaviors for the diseased and non-diseased subjects separately. With normal random effects and error terms, $\bm{Y}_{i}|D_{i}$ follows a multivariate normal distribution 
\begin{align}
\bm{Y}_{i}|D_{i} \sim MVN(\bm{\mu}_{d}, \bm{\Gamma}_{d})
\end{align}
where the means vectors and the covariance matrices are both functions of the covariates. The linear mixed models can be easily estimated from  standard statistical software packages. Then the probability $P(D_{i}=1|\bm{Y}_{i})$ can be obtained using Bayes' rule
\begin{align}
\frac{P(D_{i}=1|\bm{Y}_{i})}{P(D_{i}=0|\bm{Y}_{i})} = \frac{P(\bm{Y}_{i}|D_{i}=1)}{P(\bm{Y}_{i}|D_{i}=0)}\times \frac{P(D_{i}=1)}{P(D_{i}=0)} \label{pmm-risk}
\end{align}
where $P(D_{i}=d)$ is the prior information that is often known or can be estimated from the empirical data. The likelihood ratio $P(D_{i}=1|\bm{Y}_{i})/P(D_{i}=0|\bm{Y}_{i})$ under PMM is shown to be the optimal combination of the longitudinal biomarkers, provided that $P(\bm{Y}_{i}|D_{i}=d)$ can be accurately estimated \citep{liu2014combination}.   

PMM and SREM derive the disease risk prediction as an implicit function of the biomarkers and their observation time, and can be applied more generally, for instance, to disease early detection. It has been shown that PMM is a close approximation to SREM but the converse is not true: if SREM is the true data generation model, both PMM and SREM would have similar performances; but if PMM is the truth, SREM generally results in sub-optimal risk prediction \citep{liu2014combination, han2019accounting}.

\section{Methods for ovarian cancer early detection}\label{methods}
\begin{figure}[h]
\centerline{\includegraphics[scale=0.55]{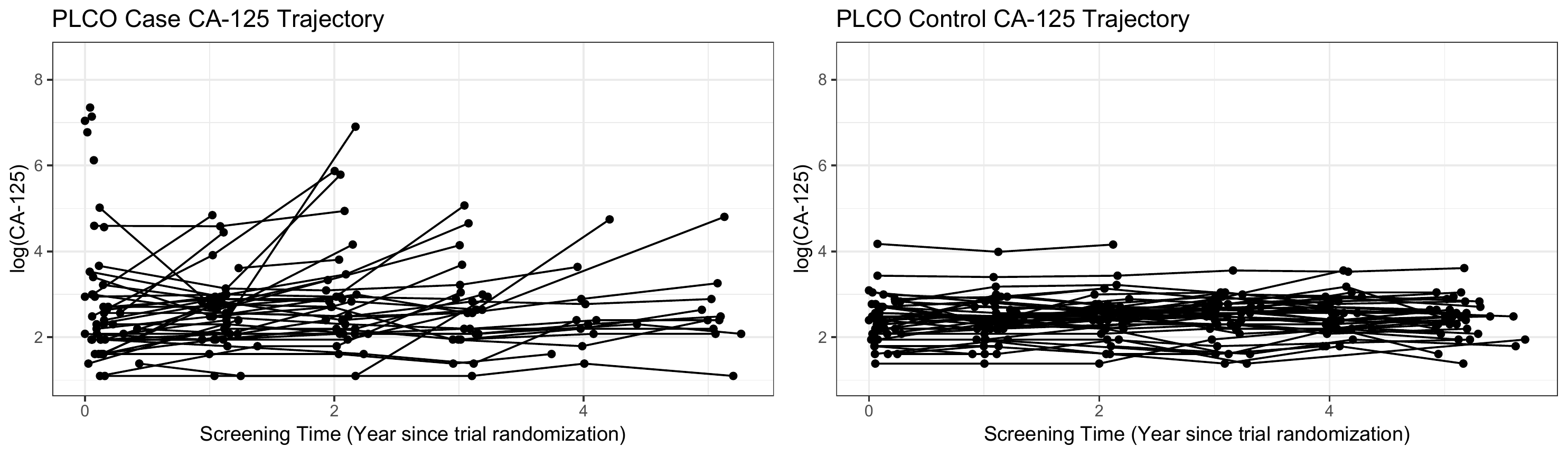}}
\caption{CA-125 trajectories of 50 cases and 50 controls that were randomly selected from the PLCO cancer screening trial \label{fig1}}
\end{figure}

Early disease detection may help to prevent death from diseases like cancer that are more curable at early stage. It is especially true for ovarian cancer, which rarely has curative treatment when detected at late stage. Good modeling of the biomarker trajectories is usually the key to accurate detection prediction. To understand the unique feature of ovarian cancer biomarker trajectories, we considered samples from the PLCO Cancer Screening Trial, where the biomarker cancer antigen 125 (CA-125) was studied for screening. Trajectories of the log-transformed CA-125 of 50 cases (women from the intervention arm of the trial who developed ovarian cancer during the screening) and 50 controls (those who did not but were also from the intervention arm of the trial so CA-125 levels were available) that were randomly selected from the PLCO Cancer Screening Trial are shown in Figure \ref{fig1}. 
The CA-125 trajectories for cases may be either flat or stay flat at first and then jump up at some time point during the screening. As a contrast, the control trajectories do not jump up and almost always keep flat. The special patterns in the case and the control trajectories require advanced modeling strategies for the CA-125 behaviors. 

In this section, we first review a well-studied approach that has been proposed for ovarian cancer early detection, namely the risk of ovarian cancer algorithm (ROCA). We then extend ROCA to identify the potential effects of the baseline age and the screening time on the biomarker trajectories and developed a maximum-likelihood approach for ROCA parameter estimation. We further propose a PMM and a SREM specifically for ovarian cancer early detection prediction. 

Under the setting of ovarian cancer early detection, $Y_{ij}$ denotes the natural logarithm transformed CA-125 measurement for the $i$th woman at time $t_{ij}$ (years since trial randomization). The time $T_{i}$ is the cancer diagnosis time for a case subject and the censored follow-up time for a control subject. 
\subsection{Risk of Ovarian Cancer Algorithm (ROCA)}
As a method specifically developed for predicting ovarian cancer early detection, ROCA separately models the longitudinal CA-125 trajectories for the cases and controls \citep{skates2001screening}. For the averaged case CA-125 trajectory, a piecewise linear model with a latent subject-specific changepoint $\tau_{i}$ conditional on the diagnosis time $T_{i}$ is assumed. The mean for cases with elevated CA-125 after $\tau_{i}$ is 
\begin{align*}
E(Y_{ij}|D_{i}=1, T_{i})=\theta_{i}+\gamma_{i}(t_{ij} - \tau_{i})^{+} 
\end{align*}
The slope of increase after $\tau_{i}$ is denoted as $\gamma_{i}$, $\theta_{i}$ is a subject-specific random intercept, and $(x)^{+}=x$ when $x>0$ and 0 otherwise. \cite{skates2001screening} assumed that the changepoint $\tau_{i}$ follows a known truncated normal distribution $N_{[T_{i}-5, T_{i}]}(T_{i}-2, 0.75^{2})$ conditional on $T_{i}$. For cases without elevated CA-125, the mean is 
\begin{align*}
E(Y_{ij}|D_{i}=1, T_{i})=\theta_{i}
\end{align*}
As for controls, a constant mean model 
\begin{align*}
E(Y_{ij}|D_{i}=0)=\theta_{i}
\end{align*}
is assumed due to the flat CA-125 behaviors. 

Parameter estimation of ROCA was implemented using the Bayesian framework. As for the ovarian cancer detection prediction, ROCA also calculates the detection probability $P(D_{i}=1|\bm{Y}_{i})$ using the Bayes' rule in (\ref{pmm-risk}) but slightly differs from PMM. The difference will be discussed in Section \ref{extend-roca}.

\subsection{Extended ROCA} \label{extend-roca}
The piecewise linear model for cases under ROCA can be rewritten as 
\begin{align}
Y_{ij} = \theta_{0} + \gamma_{0}(t_{ij} - \tau_{i})^{+} + b_{0i} + b_{1i}(t_{ij} - \tau_{i})^{+}  + \xi_{ij} \label{cs1}
\end{align}
where $\theta_{i}$ is decomposed into a constant intercept $\theta_{0}$ and a random one $b_{0i}$, and $\gamma_{i}$ is decomposed into a constant rate $\gamma_{0}$ and a random slope $b_{1i}$. Random effects $\bm{b}_{i} = (b_{0i}, b_{1i})^{\top}\sim MVN(\bm{0}, \bm{\Sigma}_{\bm{b}})$. This model depicts that the CA-125 trajectory is flat before the changepoint, and then increases with a slope of $\gamma_{0}+b_{1i}$ after the changepoint.  For case subjects without observed changepoint during the screening, the term $(t_{ij}-\tau_{i})^{+}$ is just 0, leading to a flat trajectory $Y_{ij}=\theta_{0}+b_{0i}+\xi_{ij}$. Similarly, the ROCA control model can be rewritten as a random intercept model 
\begin{align}
Y_{ij} = \theta_{0} + b_{0i} +\xi_{ij} \label{cn1}
\end{align}
For easy presentation, denote the case model (\ref{cs1}) as Case Model 1 (CS1) and the control model (\ref{cn1}) as Control Model 1 (CN1).

Several issues of ROCA requires attention: (i) the changepoint $\tau_{i}$ follows an assumed known truncated normal distribution, which may not be reasonable for all study populations; (ii) CN1 ignores the potential effects of the baseline age and the screening time on the CA-125 trajectories, inducing potentially biased inferences; (iii) the detection probability calculation may suffer loss of accuracy. Theoretically, when predicting the cancer detection of a new subject $k$, ROCA calculates the probability $P(D_{k}=1|\bm{Y}_{k})$ using the formula in (\ref{pmm-risk}). However, ROCA only obtains an approximation of $P(\bm{Y}_{k}|D_{k}=1)$, say $\tilde{P}(\bm{Y}_{k}|D_{k}=1)$, since it models $\bm{Y}_{k}|D_{k}=1, T_{k}$ rather than $\bm{Y}_{k}|D_{k}=1$. Calculating $\tilde{P}(\bm{Y}_{k}|D_{k}=1)$ requires ROCA to marginalize over the diagnosis time $T_{k}$, which is frequently unknown for the new individual $k$. To get the approximated probability $\tilde{P}(\bm{Y}_{k}|D_{k}=1)$, ROCA essentially implements the marginalization by ``borrowing'' information about the diagnosis time from participants who have already been known as cases via below formula
\begin{align*}
\tilde{P}(\bm{Y}_{k}|D_{k}=1) = \frac{1}{M}\sum_{i=1}^{M}P(\bm{Y}_{k}|D_{k}=1, T_{k}=t_{kn_{k}} + G_{i})
\end{align*}
where $M$ is the number of known cases, the gap time $G_{i}=T_{i}-t_{in_{i}}$, $T_{i}$ is the cancer diagnosis time for the $i$th known case, whose last screening time is $t_{in_{i}}$, and $t_{kn_{k}}$ is the last  screening time of the new individual $k$. This marginalization eventually leads to 
\begin{align*}
\frac{P(D_{k}=1|\bm{Y}_{k})}{P(D_{k}=0|\bm{Y}_{k})}\propto \frac{\tilde{P}(\bm{Y}_{k}|D_{k}=1)}{P(\bm{Y}_{k}|D_{k}=0)}
\end{align*} 
which may result in loss of prediction accuracy (as we observed this in simulation studies later), especially when the sample size of known cases is relatively small. 

To address the above mentioned issues, we propose to extend the original ROCA in the following ways: 
\begin{itemize}
\item[(i)] Instead of being prespecified, the parameters of the changepoint distribution are estimated, i.e., $\tau_{i}\sim N(T_{i}-\mu_{\tau}, \sigma_{\tau}^{2})$. Denote the case model (\ref{cs1}) with this unspecified changepoint distribution as Case Model 2 (CS2).
\item[(ii)] The control CA-125 trajectory is characterized by a linear mixed model that adjusts for the screening time
\begin{align}
Y_{ij} = \theta_{0}+\theta_{1}t_{ij}+b_{0i}+b_{1i}t_{ij}+\xi_{ij} \label{cn2}
\end{align}
Denote this model as Control Model 2 (CN2). 
\item[(iii)] As an alternative to (ii), the control trajectory is depicted by a linear mixed model adjusting for both the screening time and the baseline age
\begin{align}
Y_{ij} = \theta_{0}+\theta_{1}t_{ij}+\theta_{2}\text{Age}_{i}+b_{0i}+b_{1i}t_{ij}+\xi_{ij}
\end{align}
Denote this model as Control Model 3 (CN3).
\end{itemize}
Different from the Bayesian strategy in \cite{skates2001screening}, we propose a maximum-likelihood approach for the parameter estimation. The likelihood function $f(y_{ij}|t_{ij}, T_{i}, \bm{\vartheta})$ can be obtained by integrating $f(y_{ij}|t_{ij}, T_{i}, \tau_{i}, \bm{\vartheta})$ in (\ref{cs1}) over the changepoint $\tau_{i}$
\begin{align*}
f(y_{ij}|t_{ij}, T_{i}, \bm{\vartheta}) = \int f(y_{ij}|t_{ij}, T_{i}, \tau_{i}, \bm{\vartheta})g(\tau_{i}) \text{d}\tau_{i}
\end{align*}
For CS1, parameters are $\bm{\vartheta} = (\theta_{0}, \gamma_{0}, \sigma_{b_{0}}, \sigma_{b_{1}}, \rho_{b_{0}b_{1}}, \sigma_{\xi})^{\top}$ and $g(\tau_{i})$ is the PDF of $\tau_{i}$ specified by the truncated normal distribution. As for CS2, $\bm{\vartheta} = (\theta_{0}, \gamma_{0}, \sigma_{b_{0}}, \sigma_{b_{1}}, \rho_{b_{0}b_{1}}, \sigma_{\xi}, \mu_{\tau}, \sigma_{\tau})^{\top}$ and $g(\tau_{i})$ is the PDF specified in CS2. The above integration with respect to the latent changepoint could be numerically approximated by using the Gauss-Hermite quadrature. To the contrary, all of the three control models can be easily estimated from standard software packages, such as the R package {\it nlme}. Related R codes for the parameter estimation are given in the Supplementary Material.

Combinations of case and control models lead to different versions of ROCA, denoted by ROCA-CS1-CN1 (the original ROCA), ROCA-CS1-CN2, ROCA-CS1-CN3, ROCA-CS2-CN1, ROCA-CS2-CN2, and ROCA-CS2-CN3, respectively.
\subsection{PMM for Ovarian Cancer} \label{pmm-oc}
In this section, we propose a PMM such that the case model does not rely on the latent changepoint structure. More specifically, we formulated the averaged case trajectory using a linear mixed model with natural cubic splines to account for the effects of the screening time and the baseline age. The case model under PMM is given as
\begin{align}
Y_{ij} = \theta_{0} +\sum_{\ell=1}^{3}\theta_{\ell}B_{\ell}(\text{Age}_{i}, \lambda) + b_{0i}+\sum_{\ell=1}^{3}(\theta_{\ell+3}+b_{\ell i})B_{\ell}(t_{ij}, \lambda) + \xi_{ij} \label{pmm}
\end{align}
where $B_{\ell}(x, \lambda)$ is the B-spline basis of order 3 for the natural cubic spline with knot $\lambda$ decided based on $x$, $\theta_{0}$, $\theta_{\ell}$, and $\theta_{\ell+3}$ are the fixed effects, $b_{0i}$ and $b_{\ell i}$ are the random effects, and $\xi_{ij}$ is the random measurement error. For $\text{Age}_{i}$, the boundary knots were the minimum and the maximum of the baseline age of all cases, while the two internal knots were respectively set as the first and the third quantiles of all cases' baseline age. The knots for the screening time were determined in the same way. As for the averaged control trajectory, we used control models CN1, CN2, and CN3. PMM then predicts the ovarian cancer diagnosis using the formula shown in (\ref{pmm-risk}). Denote the three versions of PMM as PMM-CN1, PMM-CN2, and PMM-CN3. 

As ROCA separately models the case and the control trajectories, it can be regarded as a special case of the PMM framework. However, there are two key differences between PMM and ROCA
\begin{itemize}
\item[(i)] The first difference is that instead of using a latent changepoint structure in the case model, PMM assumes a linear mixed model with natural cubic splines and additionally adjusts for the baseline age.
\item[(ii)] The second difference lies in the calculation of the cancer diagnosis probability. As PMM directly models $\bm{Y}_{i}|D_{1}=1$, the diagnosis probability can be calculated without marginalization, avoiding the loss of prediction accuracy.
\end{itemize}
What is more, compared to ROCA, the parameter estimation under PMM could be easily implemented using standard statistical software packages rather than the Gauss-Hermite quadrature. 
\subsection{SREM for Ovarian Cancer} \label{srem-oc}
Under SREM, a linear mixed model with natural cubic splines for the screening time and the baseline age in a form same as (\ref{pmm}) was proposed to simultaneously formulate the case and the control trajectories. The knot settings of the B-spline basis for the baseline age and the screening time were determined in the same way as under PMM. For the ovarian cancer diagnosis prediction, we linked the binary outcome $D_{i}$ to the longitudinal process using a probit link function $P(D_{i}=1|\bm{b}_{i})=\Phi(\alpha_{0}+\alpha_{1}b_{i0}+\alpha_{2}b_{i1}+\alpha_{3}b_{i2}+\alpha_{4}b_{i3})$. Under this setting, the joint likelihood of $\bm{Y}_{i}$ and $D_{i}$ can be directly obtained from (\ref{srem-like}). The diagnosis probability was calculated using (\ref{srem-risk1}) with the MLEs of $\alpha_{0}$, $\alpha_{1}, \alpha_{2}, \alpha_{3}$, and $\alpha_{4}$, which were estimated by the two-stage approach described in Section \ref{SREM}.

\section{Analysis of the PLCO Ovarian Cancer Data}\label{plco}
In this section, ROCA, PMM and SREM were applied to the ovarian cancer example from the PLCO Cancer Screening Trial.
\subsection{The PLCO Ovarian Cancer Data}
The ovarian cancer dataset from the PLCO Cancer Screening Trial contains 78215 women with baseline age between 55 and 74 years at 10 screening centers across the U.S. from 1993 to 2001. Among them, 39104 women were in the intervention arm (receiving up to 6 annual screenings with CA-125 and 4 annual tests with transvaginal ultrasound) and 39111 in the control arm (under usual medical care). After the first 6 years of active screening, participants in both arms were followed for an additional 7 years \citep{skates2003calculation, buys2011effect}. Only women in the intervention arm were included in our analyses, since participants in the control arm did not receive CA-125 screening. Women who met any of the following criteria were excluded: (i) women who had historical ovarian cancer diagnosis before trial randomization; (ii) women who had bilateral oophorectomy; (iii) women who received no CA-125 screening or (iv) ovarian cancer cases who were diagnosed more than three years after the last screening test. For CA-125 screenings more than three years from the ovarian cancer diagnosis, they often have flat trajectories that are almost identical to those from controls. It is hard to tell whether any positive findings in the screening would be indicative of ovarian cancer or not \citep{pinsky2013potential}. Therefore, cases diagnosed more than three years from the last screening test were excluded from our analysis. As for the intervention arm participants chosen as our controls, their follow-up time was similarly truncated to 3 years. In addition, we removed CA-125 screening results if they were performed after the cancer diagnosis. Our analytic sample eventually included $N=30402$ women. Among them, there were $M=133$ ovarian cancer cases and 30269 controls. The median numbers of longitudinal CA-125 measurements were 4 for cases and 6 for controls due to the PLCO trial design. The ovarian cancer cases were older at the baseline, had fewer CA-125 screenings, and were more likely to have a family history of ovarian or breast cancer.  Additional descriptive statistics of the PLCO ovarian cancer samples were tabulated in Table S1 in the Supplementary Material. Pinsky et al. (2013) applied ROCA to the PLCO trial, with the aim of examining whether ROCA can result in a significant mortality benefit of screening in the intervention arm compared with the control arm \citep{pinsky2013potential}. Our analyses only used the intervention arm to compare the predictive accuracy of ROCA with PMM and SREM, in terms of the time-dependent AUC.
\subsection{Model Implementation}
Different specifications of ROCA case and control models were compared using the likelihood ratio test, Akaike information criterion (AIC) and Bayesian information criterion (BIC). Risk scores of each individual under different models were calculated using the leave-one-out cross-validation (LOOCV) technique to minimize model overfitting: longitudinal CA-125 levels of each participant were deleted from the dataset in turn, all models were estimated from the leave-one-out samples, and then the risk score of the excluded individual was calculated accordingly. Iterating this procedure across all the individuals yielded the out-of-sample risk prediction. Diagnosis prediction accuracy of the models was compared using the time-dependent AUC, during 0.5-3 years since the last CA-125 screening. The time-dependent AUC at a cutoff time $t$ is interpreted as the probability that a randomly selected ``case'', whose cancer diagnosis was before time $t$, had larger predicted risk than a randomly selected ``control'', whose cancer diagnosis was after time $t$. The 95\% confidence interval (CI) of each time-dependent AUC was calculated based on 200 bootstrapping replicates. Because of the LOOCV procedure, the nonparametric bootstrap is computationally forbidden. Therefore, instead of drawing bootstrap samples with replacement, we drew the bootstrapped parameter estimates from the fitted asymptotic multivariate normal distributions (R functions were attached in the Supplementary Material). 
\subsection{Results} \label{plco-results}
The likelihood ratio test for CS1 and CS2 showed that CS2 had better model fitting than CS1 for the PLCO cases: the negative loglikelihood values for CS1 and CS2 were 387.03 and 330.34, respectively, indicating significant difference ($p$-value: $<$ 0.0001) (AIC and BIC were in Table S2 in the Supplementary Material). The fitted parameters of CS1 and CS2 were reported in Table \ref{cs-fitting}. In CS2, $\mu_{\tau}$ and $\sigma_{\tau}$ were 1.054 years (95\% confidence interval (CI) = 0.978 to 1.130) and 0.314 year (95\% CI = 0.234 to 0.394), respectively, substantially different from the prespecified mean of 2 years and standard deviation of 0.75 year in CS1. The fitted results of PMM and SREM were in Table S3 in the Supplementary Material. 
\begin{table*}[h]
\caption{Parameter estimates for CS1 and CS2 based on 133 cases from the PLCO Cancer Screening Trial: estimate and the 95\% confidence interval (CI) were reported.  \label{cs-fitting}}
\centering
\begin{tabular*}{\textwidth}{@{\extracolsep\fill} ccc @{\extracolsep\fill}}
\toprule
\multirow{2}{*}{Parameter} & \multicolumn{1}{c}{CS1} &  \multicolumn{1}{c}{CS2} \\
\cmidrule{2-3}
 &  \multicolumn{1}{c}{Estimate (95\% CI)} &  \multicolumn{1}{c}{Estimate (95\% CI)}   \\
\midrule
$\theta_{0}$	& 2.304 (2.202, 2.406)	& 2.381 (2.281, 2.481)   \\
$\gamma_{0}$ & 1.352 (1.080, 1.624)	& 2.365 (1.922, 2.808)   \\
$\sigma_{b_{0}}$	& 0.488 (0.404, 0.572)	& 0.497 (0.423, 0.571)   \\
$\sigma_{b_{1}}$	& 1.191 (0.991, 1.391)	& 1.423 (1.102, 1.744)   \\
$\rho_{b_{0}b_{1}}$	& $-0.377$ $(-0.724, -0.030)$ 	& $-0.402$ $(-0.825, -0.021)$   \\
$\sigma_{\xi}$	& 0.265 (0.240, 0.290)	& 0.271 (0.246, 0.296)   \\
$\mu_{\tau}$		& - & 1.054 (0.978, 1.130)   \\
$\sigma_{\tau}$		& - & 0.314 (0.234, 0.394)   \\
\bottomrule
\end{tabular*}
\end{table*} 


For controls, the likelihood ratio test revealed that CN2 and CN3 were statistically better than CN1 with respective $p$-values $<$ 0.0001 and $<$ 0.0001, whereas there was no significant difference between CN2 and CN3 ($p$-value: 0.077) (see Table S2 in the Supplementary Material for details). The fitted parameters for all the three control models were reported in Table \ref{cn-fitting}, which showed that the baseline age and the screening time had small but significant effects on the CA-125 trajectories. In specific, CN3 indicated that the geometric mean level of CA-125 increased by 1.92\% (95\% CI = 1.82\% to 2.02\%) every year of follow-up, and by 0.2\% (95\% CI = 0.1\% to 0.3\%) per 1-year older in the baseline age.  
\begin{table*}[h]%
\caption{Parameter estimates for CN1, CN2, and CN3 based on 30269 controls from the PLCO Cancer Screening Trial: estimate and the 95\% confidence interval (CI) were reported. \label{cn-fitting}}
\centering
\begin{tabular*}{\textwidth}{@{\extracolsep\fill} cccc @{\extracolsep\fill}}
\toprule
\multirow{2}{*}{Parameter} & CN1 & CN2 & CN3 \\
\cmidrule{2-4}
 & Estimate (95\% CI) & Estimate (95\% CI)   & Estimate (95\% CI)   \\
\midrule
$\theta_{0}$	& 2.337 (2.332, 2.342)	& 2.296 (2.290, 2.301)	& 2.158 (2.097, 2.219)   \\
$\theta_{1}$	& -	& 0.019 (0.018, 0.020)	& 0.019 (0.018, 0.020)   \\
$\theta_{2}$		& - & -	& 0.002 (0.001, 0.003)   \\
$\sigma_{b_{0}}$	& 0.445 (0.437, 0454)	& 0.450 (0.441, 0.459)	& 0.451 (0.442, 0.460)   \\
$\sigma_{b_{1}}$	& -	& 0.039 (0.009, 0.069)	& 0.039 (0.008, 0.069)   \\
$\rho_{b_{0}b_{1}}$	& -	& $-0.141$ $(-0.181, -0.101)$	& $-0.147$ $(-0.191, -0.103)$   \\
$\sigma_{\xi}$	& 0.229 (0.224, 0.233)	& 0.215 (0.211, 0.220)	& 0.215 (0.211, 0.220)   \\
\bottomrule
\end{tabular*}
\end{table*}

\begin{figure}[h]
\centerline{\includegraphics[width=\textwidth, height=0.6\textwidth]{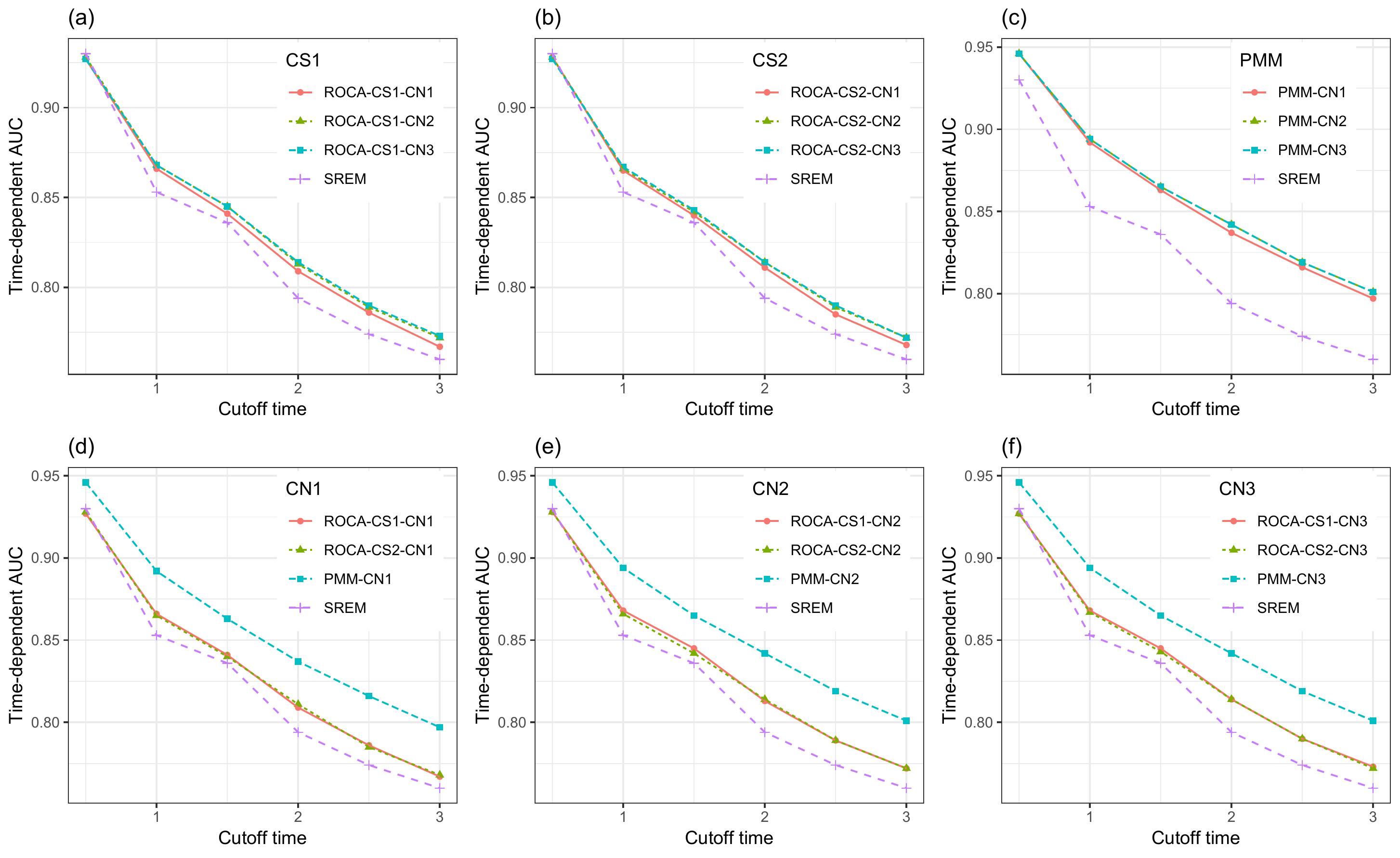}}
\caption{Time-dependent AUC comparisons for ROCA, PMM, and SREM: comparisons under the same case model setting were in figure (a)-(c) while comparisons under the same control model setting were in figure (d)-(f). \label{fig2}}
\end{figure}

\begin{figure}[h]
\centerline{\includegraphics[width=\textwidth, height=0.7\textwidth]{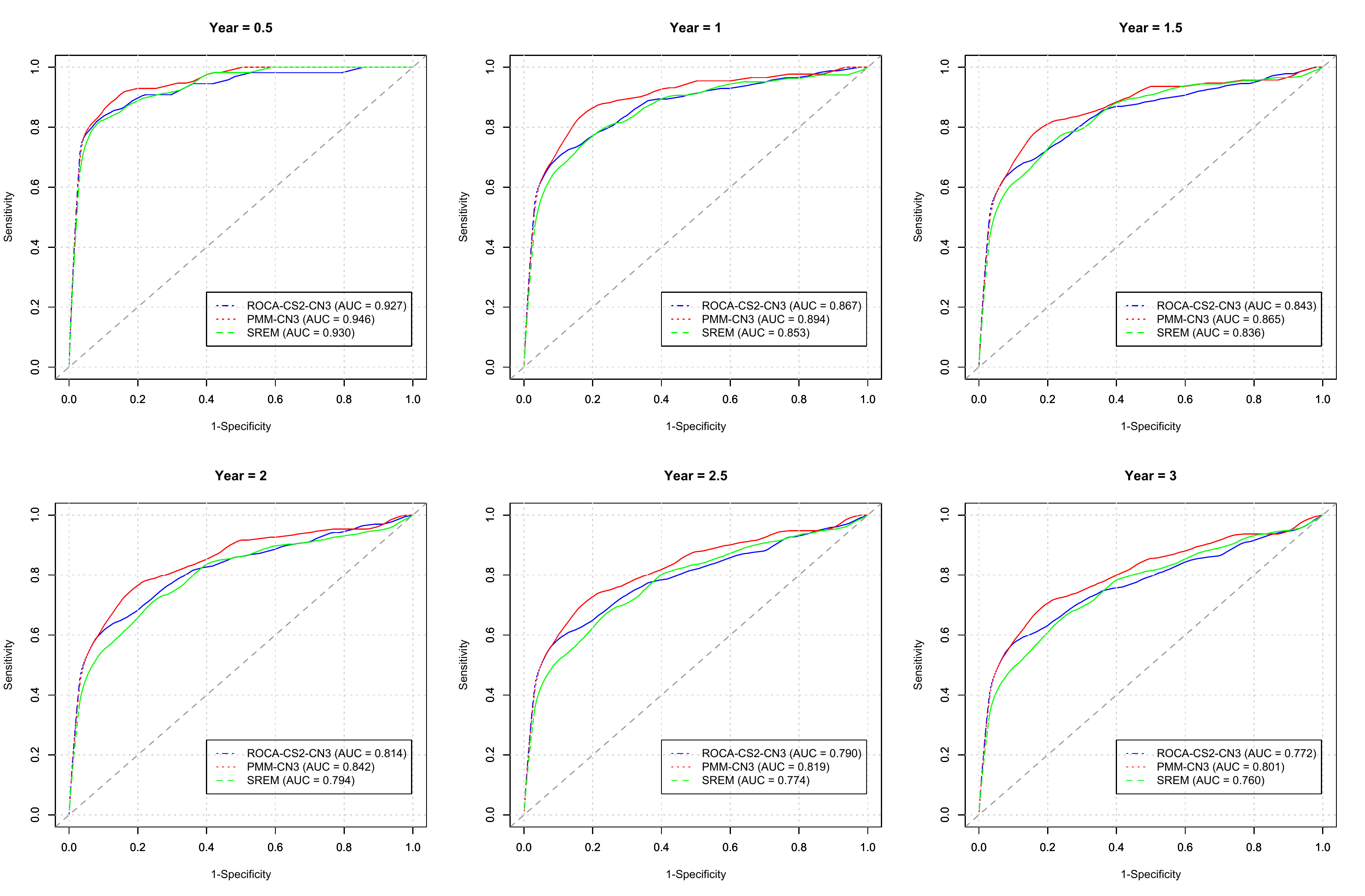}}
\caption{Time-dependent ROC curve comparisons for ROCA, PMM, and SREM across all six cutoff time. Only the best ROCA (ROCA-CS2-CN3), the best PMM (PMM-CN3), and SREM were considered. \label{fig3}}
\end{figure}

The comparisons of ROCA, PMM, and SREM with respect to their discrimination abilities were shown in Table \ref{tab3}: PMM had the highest time-dependent AUCs: 1.8-3.4\% higher than ROCA and 1.6-4.8\% higher than SREM across all six cutoff times, while SREM had the lowest AUCs. The comparison using  bootstrapping replicates further showed that PMM had significantly larger AUCs than ROCA and SREM, whereas there was no significant difference between SREM and ROCA (details were in the Supplementary Material). Among different versions of ROCA, more complex case and control models only slightly improved the time-dependent AUCs at nearly all examined cutoff time points, despite having much better goodness of fit than the original ROCA. The comparisons over all of the methods regarding the same setting of case or control model were demonstrated in Figure \ref{fig2}. Figure \ref{fig2}(a)-\ref{fig2}(c) illustrated that the improvement in the control model fitting slightly increased the diagnosis prediction accuracies of ROCA and PMM at almost all cutoff times. For example, at year 2, the AUC of ROCA-CS1-CN1 was 0.809 (95\% CI (0.797, 0.818)), compared to 0.814 (0.802, 0.827) for ROCA-CS1-CN3 that was with a more complicated control model. Figure \ref{fig2}(d)-\ref{fig2}(f) showed that more complex case model barely improved the prediction accuracy of ROCA at all cutoff times. For instance, at year 2, the AUC of ROCA-CS2-CN1 was 0.811 (0.798, 0.817), almost identical to the one of ROCA-CS1-CN1. The advantage of PMM over ROCA and SREM was  displayed in Figure \ref{fig2}(d)-\ref{fig2}(f). The AUCs of SREM were very close to those of ROCA at the beginning of the follow-up period but diminished thereafter. In addition, we compared the time-dependent ROC curves of the best ROCA (ROCA-CS2-CN3), the best PMM (PMM-CN3), and SREM across all six cutoff times. As shown in Figure \ref{fig3}, the ROC curve comparison supported the conclusion that PMM was better than ROCA and SREM, and there was no clear advantage of ROCA over SREM, though the AUCs of ROCA were larger than those of SREM. 

\begin{table*}[h]
\caption{Time-dependent AUCs of ROCA, PMM, and SREM on analyzing the ovarian cancer data from the PLCO Cancer Screening Trial. The 95\% bootstrapped confidence intervals were provided.}%
\label{tab3}
\begin{tabular*}{\textwidth}{@{\extracolsep\fill} l ccc @{\extracolsep\fill}}%
\toprule
\multirow{2}{*}{Method} & \multicolumn{3}{c}{Time-dependent AUC (95\% bootstrapped confidence interval)} \\
\cmidrule{2-4}
& Year 0.5 & Year 1.0 & Year 1.5   \\
\midrule
ROCA-CS1-CN1	& 0.927 (0.916, 0.933)	& 0.866 (0.855, 0.873)	& 0.841 (0.827, 0.850)	\\
ROCA-CS1-CN2	& 0.928 (0.917, 0.933)	& 0.868 (0.857, 0.875)	& 0.845 (0.830, 0.852)	\\
ROCA-CS1-CN3	& 0.927 (0.916, 0.934)	& 0.868 (0.857, 0.876)	& 0.845 (0.829, 0.852)	\\
ROCA-CS2-CN1	& 0.928 (0.920, 0.935)	& 0.865 (0.850, 0.869)	& 0.840 (0.824, 0.846)	\\
ROCA-CS2-CN2	& 0.928 (0.919, 0.934)	& 0.866 (0.851, 0.871)	& 0.842 (0.827, 0.850)	\\
ROCA-CS2-CN3	& 0.927 (0.918, 0.934)	& 0.867 (0.851, 0.872)	& 0.843 (0.827, 0.849)	\\
PMM-CN1 	& 0.946 (0.937, 0.953)	& 0.892 (0.887, 0.900)	& 0.863 (0.857, 0.871)	\\
PMM-CN2 	& 0.946 (0.937, 0.954)	& 0.894 (0.886, 0.902)	& 0.865 (0.857, 0.872)	\\
PMM-CN3	 & 0.946 (0.937, 0.954)	& 0.894 (0.886, 0.902)	& 0.865 (0.858, 0.872)	\\
SREM	& 0.930 (0.920, 0.938)	& 0.853 (0.843, 0.862)	& 0.836 (0.827, 0.844)	\\
\midrule
\multirow{1}{*}{Method} & Year 2.0 & Year 2.5 & Year 3.0   \\
\midrule
ROCA-CS1-CN1	& 0.809 (0.797, 0.818)	& 0.786 (0.770, 0.796)	& 0.767 (0.750, 0.773)   \\
ROCA-CS1-CN2	& 0.813 (0.801, 0.826)	& 0.789 (0.774, 0.798)	& 0.772 (0.755, 0.777)   \\
ROCA-CS1-CN3	& 0.814 (0.802, 0.827)	& 0.790 (0.774, 0.800)	& 0.773 (0.755, 0.778)   \\
ROCA-CS2-CN1	& 0.811 (0.798, 0.817)	& 0.785 (0.770, 0.796)	& 0.768 (0.750, 0.773)   \\
ROCA-CS2-CN2	& 0.814 (0.805, 0.824)	& 0.789 (0.773, 0.800)	& 0.772 (0.757, 0.779)   \\
ROCA-CS2-CN3	& 0.814 (0.805, 0.824)	& 0.790 (0.773, 0.800)	& 0.772 (0.757, 0.780)   \\
PMM-CN1 	& 0.837 (0.831, 0.844)	& 0.816 (0.807, 0.824)	& 0.797 (0.789, 0.808)   \\
PMM-CN2 	& 0.842 (0.832, 0.851)	& 0.819 (0.810, 0.828)	& 0.801 (0.791, 0.809)   \\
PMM-CN3	        & 0.842 (0.832, 0.851)	& 0.819 (0.810, 0.828)	& 0.801 (0.791, 0.809)   \\
SREM	        & 0.794 (0.787, 0.801)	& 0.774 (0.765, 0.782)	& 0.760 (0.751, 0.768)   \\
\bottomrule
\end{tabular*}
\end{table*}

\section{Simulation Studies}\label{simulation}
\subsection{Simulation Settings}
Predictive performances of ROCA, PMM and SREM were further compared in simulation studies. In order to apply ROCA, both longitudinal and survival information for cases and controls should be simulated. However, as discussed in Section \ref{pmm-oc} and Section \ref{srem-oc}, PMM and SREM predicted the ovarian cancer diagnosis using CA-125 levels directly. No explicit dependence relationship between the longitudinal process and the diagnosis time was set up under PMM and SREM, indicating the diagnosis time simulated from PMM and SREM would not be informative for the longitudinal observations, and hence can not be used to fit ROCA. To the contrary, the longitudinal process and the diagnosis time were explicitly linked together under ROCA. 

Two simulation scenarios were considered: Scenario 1 used ROCA-CS2-CN3 as the true model, whereas Scenario 2 used PMM-CN3 as the true model. We did not pursue the scenario that SREM is the true data generation model based on the results in Section \ref{plco-results} that PMM was superior to SREM. For each simulation, a training dataset of controls and cases were generated according to the true model. The controls and cases in the training dataset were used to fit the corresponding control and case models, respectively. Then the estimated models were applied to a separately generated testing dataset, which also contained the same number of controls and cases as the training dataset. The time-dependent AUCs from 0.5 to 3 years after the last CA-125 observation were calculated in the testing sample. In addition, three screening frequencies were considered in the simulations: annual, biannual and quarterly screening, aiming to examine how the performances of the above models would be affected by the frequency of CA-125 screening. 

To closely mimic the real PLCO ovarian cancer data, particularly the gap time between the diagnosis time and last screening test associated with each subject, we proposed the following data generation procedure. Let $T_{i}^{\ast}$ be the unobserved time of the ovarian cancer diagnosis, and $C_{i}$ the censoring time. The observed survival time is given by $(T_{i}, S_{i})$, where $T_{i} =\min(T_{i}^{\ast}, C_{i})$ and $S_{i}=I(T_{i}^{\ast} < C_{i})$ with $I(\cdot)$ being the indicator function. Let $n_{i}$ be the cluster size and $Y_{i1}, \ldots, Y_{in_{i}}$ be the log-transformed CA-125 marker values observed at times $t_{i1}, \ldots, t_{in_{i}}$, respectively. Let $G_{i} = T_{i}-t_{in_{i}}$ be the gap time between the last observation and the end of follow-up. The procedure to generate survival and longitudinal data is given as follow 
\begin{enumerate}
\item[1)] Step 1: simulate the survival data. The distribution of $T_{i}^{\ast}$ is simulated from an exponential distribution with rate $6\times 10^{-4}$. The distribution of $C_{i}$ is simulated from a mixture of two lognormal distributions $1/3LN(1.7, 0.4^{2})+2/3LN(2.1, 0.16^{2})$. All parameter values of the above distributions were obtained by fitting the real PLCO ovarian cancer data. The censoring time $C_{i}$ is truncated at 8.9 years as the maximum follow-up used in this analysis. Results from the PLCO ovarian cancer data showed that the Kaplan-Meier survival curves were very close to the above fitted parametric distributions for $T_{i}^{\ast}$ and $C_{i}$ (see Figure S6 in the Supplementary Material for detail). 
\item[2)] Step 2: simulate the gap time $G_{i}$. The gap time $G_{i}$ is bounded in $[t_{L}, t_{U}]$, where $t_{L}=\max(0, T_{i}-6)$ and $t_{U}=\min(3, T_{i})$. This setting is to guarantee that all the observation times $t_{i1}, \ldots, t_{in_{i}}$ would be between year 0 and year 6. For each subject $i$, we randomly sample one $G_{i}$ of the participants from the PLCO cancer data that are bounded in $[t_{L}, t_{U}]$. Meanwhile, the associated age of $G_{i}$ is also chosen as the baseline age of the $i$th subject.  
\item[3)] Step 3: simulate the cluster size $n_{i}$ and the screening time $t_{i1}, \ldots, t_{in_{i}}$. Set $n_{i} = \lfloor T_{i}-G_{i}\rfloor$, where $\lfloor x\rfloor$ denotes the maximum integer not exceeding $x$. Under the annual screening setting, the screening time $t_{ij}=0, 1, \ldots, n_{i}-1$, so that the biomarker is screened annually from year 0 to year $n_{i}-1$. The cluster size and the screening time under the settings of biannual and quarterly screening can be set similarly. In specific, under the biannual screening, the size is $2n_{i}-1$ and the time is $t_{ij}=0, 0.5, \ldots, n_{i}-1$, while they are $4n_{i}-3$ and $t_{ij}=0,0.25, 0.5, \ldots, n_{i}- 1$ under the quarterly screening. 
\item[4)] Step 4: simulate the log-transformed CA-125 values using ROCA-CS2-CN3 or PMM-CS3 based on the above simulated screening time, the diagnosis time, the event status, and the baseline age.
\end{enumerate}
Under Scenario 1, both training and testing datasets were set to contain 1000 controls and 500 cases. While under Scenario 2, the total number of controls and cases was 30402 in both the training and testing datasets. Details about the simulation setting and an R function were provided in the Supplementary Material.

Simulation results under Scenario 1 were reported in Table \ref{tab-simu1}. When the annual screening scheme was simulated (same as the PLCO ovarian cancer data), the differences among all ROCAs were small, but in general ROCA-CS2-CN3 had the best performance regarding diagnosis prediction, almost identically followed by ROCA-CS2-CN2. ROCA outperformed PMM and SREM over all cutoff time points, while SREM had the least satisfactory performance. As the number of CA-125 screenings increased, the predictive advantage of ROCA over PMM and SREM became more prominent. This is because more data points near the CA-125 changepoint became available to precisely estimate the latent changepoint structure of ROCA. To the contrary, the predictive performances of SREM deteriorated as the number of the screened CA-125 measurements increased, as the difference between case and control trajectories became evident. The discriminative accuracy of PMM did not change much. 

When PMM-CN3 was the true model, Table \ref{tab-simu2} showed that PMM outperformed ROCA and SREM. For both ROCA and PMM, complicated case or control models only provided a small amount of improvements to the time-dependent AUCs. SREM still had the least satisfactory performances. As the number of CA-125 screenings increased, all methods had improved values for the time-dependent AUC.
\section{Discussion} \label{discussion}
{\color{black}In this paper, we focused on the problem of predicting disease early detection using longitudinal biomarker measurements. Two general disease risk prediction frameworks, the shared random effects model (SREM) \citep{albert2012linear} and the pattern mixture model (PMM) \citep{liu2014combination} were considered. We showed that SREM and PMM can be applied to disease early detection in a general setting, though they were developed in a very different situation of disease risk prediction. We examined and evaluated the utility of SREM and PMM for disease early detection through an application to the early detection of ovarian cancer from the Prostate, Lung, Colorectal, and Ovarian (PLCO) Cancer Screening Trial. The predictive performances of SREM and PMM were compared with the risk of ovarian cancer algorithm (ROCA), which is specifically proposed for ovarian cancer early detection \citep{skates2001screening}. Specific formulations of SREM and PMM for predicting ovarian cancer early detection were provided. We also extended ROCA by estimating the latent changepoint structure and considering the effects of the screening time and the baseline age on the development of the biomarker trajectory. The predictive performances of the above three methods were assessed using the time-dependent AUC \citep{heagerty2000time}, such that the censored cancer diagnosis time information can be incorporated into the AUC calculation. We additionally studied the effects of three biomarker screening frequencies (annual, biannual, and quarterly) on model prediction accuracy via simulations.}

In the PLCO ovarian cancer data analysis, we found that PMM significantly outperformed ROCA and SREM. Though ROCA had slightly larger AUC values than SREM, it did not significantly differ from SREM. We noticed that a design of the PLCO trial may affect the AUC values we presented: in the intervention arm, CA-125 was used to manage women, i.e., they were referred to diagnostic evaluation when CA-125 was elevated. Therefore, the AUC based on CA-125 may be overestimated under this setting. However, such design would not affect the comparison pattern of the above approaches. The comparison is interesting as ROCA is more biologically sensible for modeling the ``jump-up'' pattern in the case marker trajectories shown in Figure \ref{fig1}. One explanation is the way that ROCA implements the prediction. ROCA models the marker profiles using unobserved changepoints conditioning on the cancer diagnosis time, which is unknown when it comes to predict the cancer onset for a new individual. To calculate the detection probability, ROCA needs to estimate the joint distribution of the longitudinal marker profiles by marginalizing out the diagnosis time. However, this marginalization may result in loss of prediction accuracy, especially when the sample size of the cases is relatively small. In addition, the estimation of the latent changepoint structure may suffer from sparse measurements around the changepoint under the annual screening design of CA-125 in the PLCO trial. We found from the simulation studies that the performance of ROCA could be substantially improved with more frequent screening data. To the contrary, SREM and PMM directly estimated the CA-125 distribution independently from the cancer diagnosis time and used natural splines to model the nonlinear marker  trajectories, avoiding the difficulty in estimating the latent changepoint. The changepoint pattern also exists in other cancer studies, for example, prostate cancer, \citep{barry2001prostate}, where the level of the biomarker prostate-specfic-antigen would be elevated before a prostate cancer case is diagnosed. This hence indicates the general applicability of SREM and PMM to disease early detection.

Extensions to the case and the control models in the original ROCA were proposed. We found that these extensions resulted in better model fitting, but only slightly improved the predictive performance of ROCA, both in the PLCO data analysis and in simulation studies. There may be several explanations of this result. First, as a rank-based measure, AUC is difficult to improve, unless the rankings of the calculated risk differ dramatically. Second, the latent changepoint structure is hard to estimate precisely with only few observations around the changepoint, and hence extending the case model with flexible changepoint distribution may not substantially change the risk calculation. Third, as the screening time and the baseline age only have small effects on the longitudinal CA-125 trajectory, incorporating them in the control model may not strongly affect the risk calculation either. 

The performance of SREM was not as good as PMM or ROCA, possibly due to that SREM models the CA-125 trajectories of both cases and controls simultaneously. This simultaneous modeling may not be a sensible choice, especially when the case trajectories are evidently different from the control ones, as shown by the ROCA simulation results  in Table \ref{tab-simu1} under the setting of quarterly screening. Furthermore, SREM may need to formulate the shared random effects and the outcome in a more complicated way rather than using a simple linear relation, calling for future methodological development.

In our study, the comparison on the discriminative performances of SREM, PMM, and ROCA on predicting the ovarian cancer early detection was based on a single biomarker CA-125, which was annually screened in the PLCO Cancer Screening Trial. Several biomarkers have been recently reported for the early detection of ovarian cancer, and studies show that incorporating those biomarkers may help to gain better prediction accuracy \citep{zhang2004three, russell2017novel, visintin2008diagnostic}. For example, \cite{russell2017novel} propose a risk prediction method for ovarian cancer by adopting three additional biomarkers together with CA-125 and demonstrate that their method has better discriminative performance than ROCA. As ROCA models CA-125 only, it cannot handle multiple biomarkers. To the contrary, as general frameworks for disease risk prediction of longitudinal studies, PMM and SREM can be easily extended to deal with studies that are with multiple biomarkers \citep{liu2014combination, zhang2012predicting}, resulting in a possible solution to predict the early detection of ovarian cancer using the recently reported markers. However, using multiple longitudinal biomarkers can be computationally challenging and requires further research.

ROCA, PMM and SREM were all constructed with the binary outcome (cancer and non-cancer) but did not fully utilize the cancer diagnosis time. Therefore, the risk calculation cannot provide an absolute risk estimation, i.e., $t$-year cancer-free survival since the last CA-125 screening. Our future investigations will focus on the extensions of the above mentioned methods to handle the survival outcome. 

In conclusion, our study shows that SREM and PMM can be applied to disease early detection in the general setting of longitudinal studies, though they were originally developed for disease risk prediction. Analysis of the ovarian cancer data from the PLCO Cancer Screening Trial finds that using PMM to predict the early detection of ovarian cancer under an annual screening setting significantly outperforms ROCA and SREM. The proposed extensions to the case and the control models in the original ROCA can significantly improve the model fitting but not necessarily the prediction accuracy. The early detection prediction accuracy of ROCA could be improved with more frequent CA-125 screenings, as the latent changepoint structure would be better estimated accordingly. 

\section*{Acknowledgments}
This study was supported by the Intramural Research Program of the National Cancer Institute, the National Institutes of Health (NIH), United States. This work utilized the computational resources of the NIH High-Performance Computing Biowulf cluster (http://hpc.nih.gov).

\bibliographystyle{apalike} 
\bibliography{manuscript_arXiv_Hanetal2019}

\begin{center}
\begin{sidewaystable*}[h]%
\caption{Expected time-dependent AUCs and the associated standard deviations (SDs) of ROCA, PMM, and SREM under Scenario 1 that ROCA-CS2-CN3 is the true model. Annual, biannual, and quarterly screening frequencies were used.\label{tab-simu1}}
\centering
\begin{tabular*}{\textwidth}{@{\extracolsep\fill} ll cccccc @{\extracolsep\fill}}
\toprule
Screening & \multirow{2}{*}{Method} & \multicolumn{6}{c}{Expected time-dependent AUC (SD in \%)} \\
\cmidrule{3-8}
frequency & & Year 0.5 & Year 1.0 & Year 1.5 & Year 2.0 & Year 2.5 & Year 3.0   \\
\midrule
\multirow{10}{*}{Annual}	
& ROCA-CS1-CN1	& 0.915 (2.1)	& 0.858 (2.1)	& 0.785 (1.9)	& 0.738 (1.8)	& 0.710 (1.7)	& 0.698 (1.5)   \\
& ROCA-CS1-CN2	& 0.913 (2.1)	& 0.856 (2.2)	& 0.787 (2.0)	& 0.744 (1.8)	& 0.720 (1.6)	& 0.711 (1.5)   \\
& ROCA-CS1-CN3	& 0.913 (2.1)	& 0.856 (2.2)	& 0.787 (2.0)	& 0.744 (1.8)	& 0.720 (1.6)	& 0.711 (1.5)   \\
& ROCA-CS2-CN1	& 0.916 (2.0)	& 0.859 (2.0)	& 0.786 (1.9)	& 0.739 (1.8)	& 0.712 (1.7)	& 0.700 (1.5)   \\
& ROCA-CS2-CN2	& 0.911 (2.1)	& 0.853 (2.2)	& 0.785 (2.0)	& 0.744 (1.8)	& 0.721 (1.7)	& 0.712 (1.5)   \\
& ROCA-CS2-CN3	& 0.911 (2.1)	& 0.853 (2.2)	& 0.785 (2.0)	& 0.744 (1.8)	& 0.721 (1.7)	& 0.712 (1.5)   \\
& PMM-CN1	& 0.906 (2.3)	& 0.841 (2.2)	& 0.768 (1.9)	& 0.722 (1.7)	& 0.695 (1.6)	& 0.682 (1.4)   \\
& PMM-CN2	& 0.903 (2.3)	& 0.839 (2.2)	& 0.768 (2.0)	& 0.724 (1.7)	& 0.700 (1.6)	& 0.689 (1.4)   \\
& PMM-CN3	& 0.903 (2.3)	& 0.839 (2.2)	& 0.768 (2.0)	& 0.724 (1.7)	& 0.700 (1.6)	& 0.689 (1.4)   \\
& SREM	& 0.881 (2.6)	& 0.811 (2.1)	& 0.726 (2.0)	& 0.670 (2.0)	& 0.635 (1.8)	& 0.613 (1.7)   \\
\midrule
\multirow{10}{*}{Biannual}	
& ROCA-CS1-CN1	& 0.914 (2.3)	& 0.864 (2.0)	& 0.792 (2.1)	& 0.747 (2.0)	& 0.724 (1.7)	& 0.712 (1.6)   \\
& ROCA-CS1-CN2	& 0.912 (2.3)	& 0.864 (2.0)	& 0.798 (2.0)	& 0.758 (2.0)	& 0.738 (1.6)	& 0.729 (1.5)   \\
& ROCA-CS1-CN3	& 0.912 (2.3)	& 0.864 (2.0)	& 0.798 (2.0)	& 0.758 (2.0)	& 0.738 (1.6)	& 0.729 (1.5)   \\
& ROCA-CS2-CN1	& 0.914 (2.3)	& 0.865 (2.0)	& 0.795 (2.1)	& 0.750 (2.0)	& 0.727 (1.7)	& 0.716 (1.6)   \\
& ROCA-CS2-CN2	& 0.910 (2.3)	& 0.862 (1.9)	& 0.797 (2.0)	& 0.758 (2.0)	& 0.739 (1.7)	& 0.730 (1.5)   \\
& ROCA-CS2-CN3	& 0.910 (2.3)	& 0.862 (1.9)	& 0.797 (2.0)	& 0.758 (2.0)	& 0.739 (1.7)	& 0.730 (1.5)   \\
& PMM-CN1	& 0.898 (2.6)	& 0.835 (2.2)	& 0.756 (2.3)	& 0.705 (1.9)	& 0.678 (1.7)	& 0.661 (1.6)   \\
& PMM-CN2	& 0.896 (2.5)	& 0.835 (2.2)	& 0.761 (2.3)	& 0.715 (2.0)	& 0.690 (1.7)	& 0.675 (1.6)   \\
& PMM-CN3	& 0.896 (2.5)	& 0.835 (2.2)	& 0.761 (2.3)	& 0.715 (2.0)	& 0.690 (1.7)	& 0.675 (1.6)   \\
& SREM	& 0.865 (2.8)	& 0.792 (2.6)	& 0.709 (2.2)	& 0.658 (2.0)	& 0.625 (1.8)	& 0.604 (1.8)   \\
\midrule
\multirow{10}{*}{Quarterly}	
& ROCA-CS1-CN1	& 0.919 (2.0)	& 0.867 (1.8)	& 0.800 (1.7)	& 0.761 (1.8)	& 0.741 (1.9)	& 0.734 (1.8)   \\
& ROCA-CS1-CN2	& 0.918 (2.0)	& 0.870 (1.9)	& 0.810 (1.8)	& 0.778 (1.8)	& 0.764 (1.7)	& 0.762 (1.8)   \\
& ROCA-CS1-CN3	& 0.918 (2.0)	& 0.870 (1.9)	& 0.810 (1.8)	& 0.778 (1.8)	& 0.764 (1.7)	& 0.762 (1.8)   \\
& ROCA-CS2-CN1	& 0.919 (2.0)	& 0.869 (1.8)	& 0.803 (1.7)	& 0.766 (1.8)	& 0.747 (1.8)	& 0.741 (1.8)   \\
& ROCA-CS2-CN2	& 0.917 (2.1)	& 0.870 (2.0)	& 0.811 (1.9)	& 0.780 (1.8)	& 0.767 (1.7)	& 0.765 (1.8)   \\
& ROCA-CS2-CN3	& 0.917 (2.1)	& 0.870 (2.0)	& 0.811 (1.9)	& 0.780 (1.8)	& 0.767 (1.7)	& 0.765 (1.8)   \\
& PMM-CN1	& 0.900 (2.5)	& 0.832 (2.3)	& 0.750 (2.0)	& 0.701 (2.0)	& 0.674 (2.0)	& 0.658 (1.9)   \\
& PMM-CN2	& 0.900 (2.5)	& 0.837 (2.2)	& 0.761 (1.9)	& 0.718 (1.9)	& 0.695 (2.0)	& 0.683 (1.9)   \\
& PMM-CN3	& 0.900 (2.5)	& 0.837 (2.2)	& 0.761 (1.9)	& 0.718 (1.9)	& 0.695 (2.0)	& 0.683 (1.9)   \\
& SREM	& 0.864 (3.0)	& 0.785 (2.5)	& 0.699 (2.2)	& 0.647 (2.0)	& 0.614 (1.8)	& 0.593 (1.8)   \\
\bottomrule
\end{tabular*}
\end{sidewaystable*}
\end{center}
\begin{center}
\begin{sidewaystable*}[h]%
\caption{Expected time-dependent AUCs and the associated standard deviation (SDs) of ROCA, PMM, and SREM under Scenario 2 that PMM-CN3 is the true model. Annual, biannual, and quarterly screening frequencies were used. \label{tab-simu2}}
\centering
\begin{tabular*}{\textwidth}{@{\extracolsep\fill} ll cccccc @{\extracolsep\fill}}
\toprule
Screening & \multirow{2}{*}{Method} & \multicolumn{6}{c}{Expected time-dependent AUC (SD in \%)} \\
\cmidrule{3-8}
frequency & & Year 0.5 & Year 1.0 & Year 1.5 & Year 2.0 & Year 2.5 & Year 3.0   \\
\midrule
\multirow{10}{*}{Annual}	
& ROCA-CS1-CN1	& 0.929 (2.0)	& 0.913 (1.7)	& 0.907 (1.6)	& 0.885 (1.6)	& 0.871 (1.7)	& 0.855 (1.6)   \\
& ROCA-CS1-CN2	& 0.933 (1.9)	& 0.917 (1.6)	& 0.912 (1.6)	& 0.891 (1.6)	& 0.879 (1.8)	& 0.864 (1.6)   \\
& ROCA-CS1-CN3	& 0.932 (1.9)	& 0.917 (1.6)	& 0.912 (1.6)	& 0.891 (1.6)	& 0.879 (1.8)	& 0.864 (1.6)   \\
& ROCA-CS2-CN1	& 0.929 (2.0)	& 0.913 (1.7)	& 0.907 (1.7)	& 0.885 (1.7)	& 0.871 (1.8)	& 0.855 (1.7)   \\
& ROCA-CS2-CN2	& 0.932 (1.9)	& 0.916 (1.7)	& 0.911 (1.7)	& 0.890 (1.7)	& 0.878 (1.8)	& 0.863 (1.7)   \\
& ROCA-CS2-CN3	& 0.932 (1.9)	& 0.916 (1.7)	& 0.911 (1.7)	& 0.890 (1.7)	& 0.878 (1.8)	& 0.863 (1.7)   \\
& PMM-CN1	& 0.948 (1.5)	& 0.931 (1.5)	& 0.924 (1.5)	& 0.900 (1.6)	& 0.884 (1.7)	& 0.864 (1.7)   \\
& PMM-CN2	& 0.948 (1.5)	& 0.932 (1.6)	& 0.925 (1.6)	& 0.903 (1.6)	& 0.888 (1.7)	& 0.870 (1.7)   \\
& PMM-CN3	& 0.948 (1.5)	& 0.932 (1.6)	& 0.925 (1.6)	& 0.903 (1.6)	& 0.888 (1.7)	& 0.870 (1.7)   \\
& SREM	& 0.861 (2.8)	& 0.836 (2.8)	& 0.826 (2.8)	& 0.784 (2.9)	& 0.754 (3.0)	& 0.714 (2.9)   \\
\midrule
\multirow{10}{*}{Biannual}	
& ROCA-CS1-CN1	& 0.952 (1.6)	& 0.932 (1.5)	& 0.929 (1.6)	& 0.911 (1.7)	& 0.906 (2.1)	& 0.896 (2.1)   \\
& ROCA-CS1-CN2	& 0.957 (1.4)	& 0.937 (1.3)	& 0.935 (1.5)	& 0.918 (1.4)	& 0.914 (1.7)	& 0.906 (1.7)   \\
& ROCA-CS1-CN3	& 0.957 (1.5)	& 0.937 (1.4)	& 0.935 (1.5)	& 0.918 (1.4)	& 0.914 (1.7)	& 0.906 (1.7)   \\
& ROCA-CS2-CN1	& 0.953 (1.6)	& 0.932 (1.5)	& 0.930 (1.6)	& 0.912 (1.8)	& 0.907 (2.0)	& 0.897 (2.1)   \\
& ROCA-CS2-CN2	& 0.958 (1.3)	& 0.937 (1.3)	& 0.935 (1.5)	& 0.918 (1.4)	& 0.914 (1.6)	& 0.906 (1.6)   \\
& ROCA-CS2-CN3	& 0.958 (1.4)	& 0.937 (1.4)	& 0.935 (1.5)	& 0.918 (1.4)	& 0.915 (1.6)	& 0.906 (1.6)   \\
& PMM-CN1	& 0.966 (1.2)	& 0.949 (1.4)	& 0.946 (1.5)	& 0.926 (1.3)	& 0.919 (1.3)	& 0.907 (1.5)   \\
& PMM-CN2	& 0.969 (1.1)	& 0.952 (1.3)	& 0.949 (1.4)	& 0.930 (1.1)	& 0.924 (1.1)	& 0.915 (1.2)   \\
& PMM-CN3	& 0.969 (1.1)	& 0.952 (1.3)	& 0.949 (1.4)	& 0.930 (1.1)	& 0.925 (1.1)	& 0.915 (1.2)   \\
& SREM	& 0.869 (1.7)	& 0.853 (1.7)	& 0.843 (1.5)	& 0.803 (1.9)	& 0.779 (2.2)	& 0.731 (2.9)   \\
\midrule
\multirow{10}{*}{Quarterly}	
& ROCA-CS1-CN1	& 0.958 (0.6)	& 0.946 (0.9)	& 0.943 (1.0)	& 0.933 (1.0)	& 0.929 (0.9)	& 0.921 (0.8)   \\
& ROCA-CS1-CN2	& 0.964 (0.5)	& 0.954 (0.8)	& 0.951 (1.0)	& 0.943 (1.0)	& 0.940 (0.9)	& 0.934 (0.8)   \\
& ROCA-CS1-CN3	& 0.964 (0.5)	& 0.954 (0.8)	& 0.951 (1.0)	& 0.943 (1.0)	& 0.940 (1.0)	& 0.934 (0.8)   \\
& ROCA-CS2-CN1	& 0.959 (0.6)	& 0.947 (0.8)	& 0.944 (1.0)	& 0.934 (1.0)	& 0.930 (0.9)	& 0.922 (0.8)   \\
& ROCA-CS2-CN2	& 0.965 (0.5)	& 0.955 (0.8)	& 0.952 (1.0)	& 0.944 (1.0)	& 0.941 (1.0)	& 0.934 (0.9)   \\
& ROCA-CS2-CN3	& 0.965 (0.5)	& 0.954 (0.8)	& 0.952 (1.0)	& 0.944 (1.0)	& 0.940 (1.0)	& 0.934 (0.9)   \\
& PMM-CN1	& 0.965 (0.8)	& 0.956 (1.0)	& 0.952 (1.1)	& 0.940 (1.0)	& 0.934 (0.9)	& 0.922 (0.9)   \\
& PMM-CN2	& 0.970 (0.6)	& 0.961 (0.9)	& 0.958 (1.0)	& 0.949 (1.0)	& 0.944 (0.9)	& 0.934 (0.8)   \\
& PMM-CN3	& 0.970 (0.6)	& 0.961 (0.9)	& 0.958 (1.0)	& 0.948 (1.0)	& 0.944 (0.9)	& 0.934 (0.8)   \\
& SREM	& 0.892 (1.6)	& 0.859 (1.7)	& 0.854 (1.4)	& 0.802 (1.9)	& 0.784 (2.1)	& 0.749 (2.7)   \\
\bottomrule
\end{tabular*}
\end{sidewaystable*}
\end{center}
\end{document}